\documentclass[superscriptaddress,nofootinbib,preprintnumbers,rmp]{revtex4}
\usepackage{amsmath}
\usepackage{amsfonts}
\usepackage{amssymb}
\usepackage{epsfig}
\newcommand{\be}{\begin{equation}}
\newcommand{\ee}{\end{equation}}

\begin{document}

\title{Static and dynamic properties of a reversible gel}


\author{P.I. Hurtado}
\affiliation{Institute \emph{Carlos I} for Theoretical and Computational 
Physics, and 
Departamento de Electromagnetismo y F\'{\i}sica de la Materia, 
Universidad de Granada, Granada 18071, Spain}

\author{P. Chaudhuri}
\affiliation{Laboratoire des Colo\"ides, Verres et Nanomat\'eriaux, 
Universit\'e Montpellier 2 and CNRS, Montpellier 34095, France}

\author{L. Berthier}
\affiliation{Laboratoire des Colo\"ides, Verres et 
Nanomat\'eriaux, Universit\'e Montpellier 2 
and CNRS, Montpellier 34095, France}

\author{W. Kob}
\affiliation{Laboratoire des Colo\"ides, Verres et 
Nanomat\'eriaux, Universit\'e Montpellier 2
and CNRS, Montpellier 34095, France}

\begin{abstract}
We study a microscopically realistic model of a physical gel and use computer simulations to investigate its 
static and dynamic properties at thermal equilibrium. The phase diagram comprises a sol phase, 
a coexistence region ending at a critical point, a gelation line, and an 
equilibrium gel phase unrelated to phase separation. The global structure of
the gel is homogeneous, but the stress is supported by a fractal network.
Gelation results in a dramatic slowing down of the dynamics, which can be used to locate the transition, 
which otherwise shows no structural signatures.
Moreover, the equilibrium gel dynamics is highly 
heterogeneous as a result of the presence of 
particle families with different mobilities.
An analysis of gel dynamics in terms of mobile and arrested particles allows us to elucidate several differences between the dynamics 
of equilibrium gels and that of glass-formers.
\end{abstract}

\maketitle

\section{Introduction}

A gel can be roughly defined as a low-density, disordered, solid material composed by a liquid matrix in 
which dispersed particles form a very open network. In this way, a gel can be notably elastic and jellylike, 
as for instance gelatin, or rather solid and rigid, as silica 
gel~\cite{larson}. Though gels are materials common 
to everyday experience, their structural and dynamical properties remain puzzling in many respects~\cite{Zacca}.
This is mostly due to the wide window of timescales and lengthscales 
which determine their physical behavior, 
e.g. from the molecular size of particles
in the solvent to supramolecular structures~\cite{larson}.
An important parameter for a gel is the typical lifetime of the interparticle bonds which define the underlying stress-sustaining network. 
In the limit of very strong,  \emph{permanent} bonds (with typical energies much larger than $k_B T$), diffusion-limited cluster 
aggregation (DLCA) may lead to the formation of a fractal, system-spanning network usually termed as chemical gel and whose 
properties follow directly from geometry. In particular, gelation in chemical gels can be unambiguously identified with percolation.
On the other hand, 
weaker bonds (with energies competing directly with $k_B T$) 
result in the formation of physical (or reversible) gels, where the links have a finite lifetime and 
the transient character of the network results in a complex interplay between structure and dynamics, leading to non-trivial flow properties.

Despite its ubiquity, the nature of physical gelation is still under intense debate and several mechanisms have been 
proposed to account for this transition, among them 
geometric percolation~\cite{we,napoli,jullien}, the glass transition~\cite{fs1,puertas,ema}, 
and arrested phase separation~\cite{fs2,dave}. 
For instance, detailed experiments performed with colloidal particles 
with tunable interactions~\cite{colloids} revealed that a non-trivial interplay between  phase separation and kinetic arrest may 
produce gel-like structures. Associating polymers constitute another well-studied example of reversible gels~\cite{larson}. 
In that case, gels can be obtained far from  phase separation, producing viscoelastic materials with highly 
non-linear rheological properties that are not well understood~\cite{polymers,porte}. In fact, one may speak about different \emph{routes}
to gelation~\cite{Zacca}: a nonequilibrium route based on kinetic arrest during spinodal decomposition (an irreversible process), and an equilibrium 
route in which the gel state is reached from an ergodic phase always allowing for the equilibration of the underlying network structure (as in the case 
of polymer association). In all cases, the emergence of the gel phase ensues some degree of kinetic arrest (at least for low wave-vector modes) which 
reflects the appearance of a percolating macroscopic structure 
capable of slowing down particle motions over long timescales. 
This kinetic arrest implies
in many cases a close similarity between gelation and glass formation~\cite{larson}. We explain below this similarity for the case of equilibrium gels, 
but discuss also important differences.

In this paper we report 
results for a recently introduced model of a reversible physical gel which is
microscopically realistic (we are in fact inspired by one particular material), complementing the results already published in Ref. \cite{we}. 
Moreover, we describe in detail a hybrid Monte Carlo / molecular dynamics numerical approach designed to successfully bridge the gap between 
microscopic details and macroscopic observations, while offering deep insight on the nature of physical gels. 

\section{A Model of Reversible Gelation}

\begin{figure}[t]
\includegraphics[height=.2\textheight,clip]{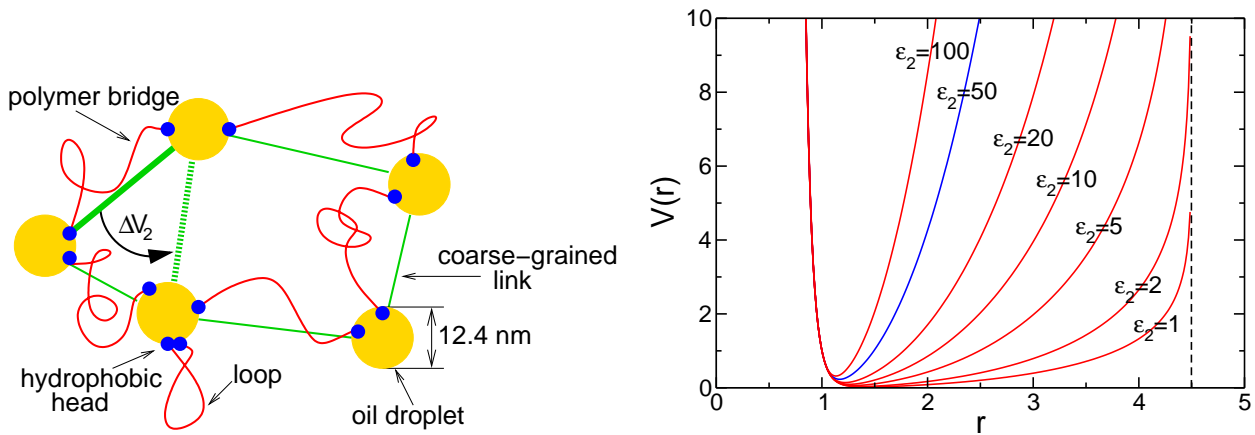}
\includegraphics[height=.24\textheight,clip]{potential.eps}
\caption{\small Left: Sketch of the oil-in-water microemulsion and our modelling strategy. Both bridging polymers 
and the coarse-grained effective links used to model them are depicted. The thicker and dashed links represent a
Monte Carlo move for the connectivity matrix.
Right: Potential energy for two droplets connected by a polymer, for different amplitudes $\epsilon_2$. 
In this paper we choose $\epsilon_2=50$ (in blue). Notice the short-range 
repulsion at $r=\sigma=1$, and the divergence at the maximum polymer 
extension, $\ell=3.5\sigma$.}
\label{sketch}
\end{figure}

Our model is inspired by a specific material \cite{porte}: A microemulsion of stable and monodisperse 
oil droplets in water mixed with telechelic polymers. These polymers are long hydrophilic chains with hydrophobic end caps.
Hydrophobicity  guarantees that a polymer can form a loop around a single droplet or, more interestingly, a bridge 
between two droplets, see the sketch in Fig. \ref{sketch} (left). Polymer bridges induce an effective entropic attraction between the droplets they connect,
due to the configurational constraint resulting from the two polymer end caps lying at a given distance.
For sufficiently high polymer and droplet concentrations, a network
of connected droplets spanning the entire system
can be formed and the system becomes a soft solid. However, thermally activated extraction of the 
hydrophobic heads leads to a slow reorganization of the network structure, and the material eventually
flows at long times: it is a transient network fluid~\cite{tanaka}. 

This material is interesting because functionality, lifetime of the bonds, 
volume fraction, strength of the links can all be adjusted independently, which is not always possible 
in attractive colloids~\cite{colloids}, or in previous model systems, unless specific adhoc assumptions are 
made~\cite{fs1,puertas,ema,fs2}. For instance, attractive colloidal gels can exhibit long-living bonds whenever a large 
attraction strength between the particles is present, but such a 
strong attraction simultaneously leads to liquid-gas phase separation
preventing the formation of a gel at thermal equilibrium~\cite{colloids}. 
On the other hand, in our model system both the bond lifetime and the attraction strength can be independently adjusted, allowing a different 
route to colloidal gelation, as discussed below.

Modelling such a complicated self-assembly of solvent, droplets and complex polymers is a  challenge because 
of the wide range of scales involved. In our simplified model we neglect the solvent (which might only affect short time dynamics) and 
include polymers and droplets as the elementary objects. Moreover, since the internal dynamics of the polymers ($\sim 10^{-8}$ s)
is much faster than the gel dynamics ($\sim 1$ s), we coarse-grain the polymer description and only retain their effect as links 
inducing an effective interaction between the two droplets they connect, see left panel in Fig. \ref{sketch}. 
Coarse-graining is a crucial step for efficient large scale 
simulations, not used in previous models~\cite{tanakasimu}.

We consider an assembly of $N$ droplets of diameter $\sigma$ and mass $m$, interacting, in the absence
of polymers, via a pair potential typical of soft spheres
\be
V_{1}(r_{ij}) = \epsilon_1 \left(\frac{\sigma}{r_{ij}} \right)^{14} \, , \nonumber
\ee
where $r_{ij}$ is the distance between droplets $i$ and $j$, and $\epsilon_1$ is an energy scale. The potential is cut off 
and regularized at a finite distance $2.5 \, \sigma$. In addition, $N_{\text{p}}$ polymers of maximal extension $\ell$ 
can form bridges between droplets, or loops. We assume 
that polymer loops have an energy cost $\epsilon_{0}$, 
but no effect on droplet dynamics. 
On the other hand, bridging polymers induce an entropic attraction between connected droplets, which we model using
the classic FENE form, which is 
obtained assuming a simple random walk description for 
polymers but retains the idea that polymers have a maximal 
possible extension~\cite{larson}:
\be
V_2(r_{ij}) = - \epsilon_2 \ln \left[ 1-\frac{(r_{ij}-\sigma)^2}{\ell^2}
  \right] \, , \qquad r_{ij} < \ell \, ,
\label{fene}
\ee 
so that polymers act as linear springs at small elongation, but cannot become
longer than $\ell$.
The right panel in Fig. \ref{sketch} shows the total interaction potential for two droplets of diameter $\sigma=1$, linked by one telechelic polymer at 
different values of $\epsilon_2$.
A system configuration is thus specified by the droplets positions and velocities, $\{{\bf r}_i(t), {\bf v}_i(t)\}$, and by the polymer 
$N \times N$ connectivity matrix, $\{C_{ij}\}$, where $C_{ij}$ is the number of polymers connecting droplets $i$ and $j$. 
The Hamiltonian is thus
\be 
{\cal H} = 
\sum_{i=1}^N \Big( \frac{m}{2} {\bf v}_i^2+  C_{ii} \epsilon_{0} + 
\sum_{j>i} \left[ V_1(r_{ij}) + C_{ij} V_2(r_{ij}) \right] \Big).
\ee

Simulation proceeds by solving Newton's equations for the droplets. Lengthscales are given in units of $\sigma$, energy in units
of $\epsilon_1$, and times in units of $\sqrt{m \sigma^2/\epsilon_1}$. We use the velocity Verlet algorithm with discretization $h=0.005$. 
Simultaneously, we use Monte Carlo dynamics to evolve the polymers. 
In one elementary move, a polymer is chosen at random, and
one of its end caps is moved to a randomly chosen neighboring droplet. This proposed move is accepted with a Metropolis rate 
\be
\tau_{\text{link}}^{-1} \, \text{min}\left[1,\text{exp}\left(-\frac{\Delta V_2}{k_B T}\right)\right] \, , \nonumber
\ee 
where $\Delta V_2$ is the potential energy change during the move, 
$T$ is the temperature, $k_B$ the Boltzmann constant, and $\tau_{\text{link}}$ 
controls the timescale for polymer rearrangements. This process is sketched in Fig. \ref{sketch} (left). 
Notice that the droplets configuration remains unchanged during a polymer move, an
approximation justified by the broad separation of time scales
between the fast polymer dynamics and the much slower droplet motion.
In experiments, $\tau_{\text{link}}$ has an Arrhenius behavior
associated to the excitation cost for polymer extraction.
We set $\ell=3.5 \, \sigma$ as measured in experiments~\cite{porte}, $T=1$,
and $\epsilon_0=1$.
We found little influence of $\epsilon_2$ on the phase
diagram, though the choice $\epsilon_2=50$ seems most appropriate for a proper
numerical comparison with experiments (see Fig. \ref{sketch}, right).
The relevant control parameters for thermodynamics are the droplet 
volume fraction, $\phi=\pi \sigma^3 N / (6V)$, where $V$ is the volume,
and the number of polymer heads per 
droplet, $R=2 N_{\text{p}}/N$. Additionally, the dynamics
is crucially affected by $\tau_{\rm link}$, which
controls the rate for polymer extraction but
has no influence on static properties. 
We performed simulations for a wide range of parameters,
$R\in[0,18]$, $\phi\in [0.01,0.3]$, $\tau_{\rm link} \in [1,10^4]$,
$N \in [10^3,10^4]$.

\section{Phase Diagram and Structure}

\begin{figure}[t]
\includegraphics[height=.25\textheight,clip]{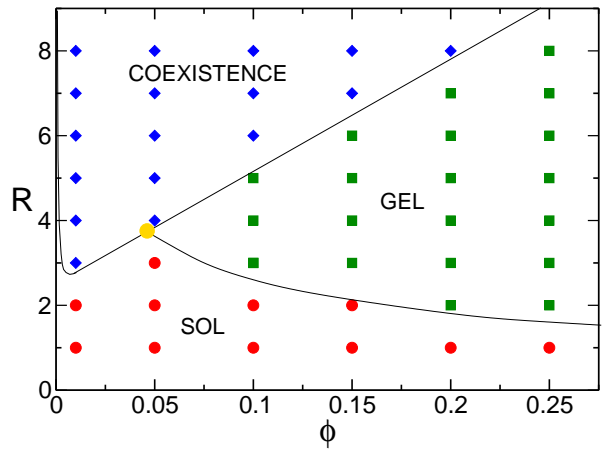}
\caption{\small  Phase diagram for a wide range of volume fractions, $\phi$, and number 
of polymer heads per droplet, $R$. Symbols are the investigated state points in the sol ($\bigcirc$), gel ($\Box$), 
and phase separated ($\diamond$) regions. A yellow point shows the 
approximate location of the critical point. Transition lines are sketched.}
\label{phase}
\end{figure}

\begin{figure}[t]
\includegraphics[height=.35\textheight,clip]{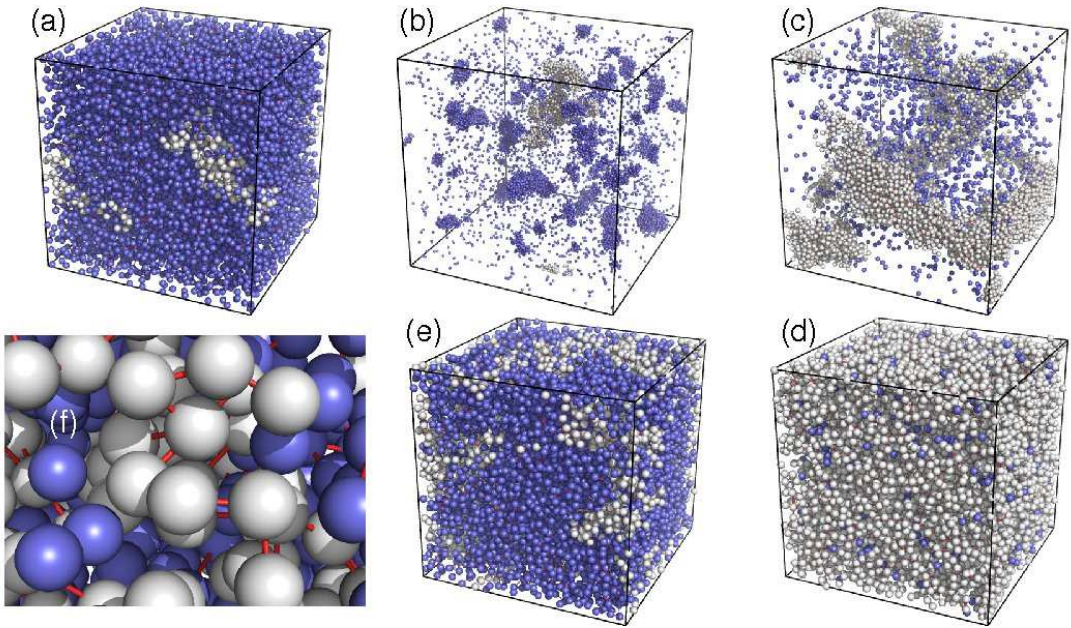}
\caption{\small Simulation snapshots for $N=10^4$. Particles colored in light gray belong to the largest cluster.
(a) Sol phase, $\phi=0.15$, $R=2$. (b) Nucleation regime, $\phi=0.01$, $R=8$. (c) Spinodal decomposition,  
$\phi=0.05$, $R=8$. (d) Structure of the physical gel
for $\phi=0.2$, $R=4$. (e) Gel right at percolation, $\phi=0.2$, $R=2$. 
A percolating (light gray) cluster of droplets connected by 
(red) polymers, through which the remaining (dark blue) droplets can diffuse, 
see zoom (f).  
}
\label{snap}
\end{figure}

The phase diagram, as obtained after a systematic exploration of the control parameter space,
is shown in Fig.~\ref{phase}. Its topology is in good agreement with experiments~\cite{porte}. It comprises a sol phase, a coexistence region ending at a critical point, a 
gelation line determined by geometric percolation, and an equilibrium gel phase.

In the low-$\phi$, low-$R$ region, the system resembles 
a dilute assembly of soft spheres, and has the properties of a simple liquid. 
This is the sol phase, see Fig.~\ref{snap}-a. 
There the largest cluster of droplets connected by polymers 
(shown in light gray in Fig.~\ref{snap}) does 
not percolate across the system. Increasing $R$ increases the effective attraction between droplets,
so that phase separation occurs at large $R$ between a low-$\phi$, low-$R$
phase and a large-$\phi$, large-$R$ phase~\cite{prlporte}, 
see Figs.~\ref{snap}-b and \ref{snap}-c. 
We detect phase coexistence from direct inspection of typical 
configurations in obvious cases. We also 
measured the static structure factor
\be
S(q) =  \left\langle \frac{1}{N}
\sum_{k=1}^N \sum_{j=1}^N \exp [ i {\bf q} \cdot ({\bf r}_j - {\bf r}_k)] 
\right\rangle \, , \nonumber
\ee
which exhibits a $q^{-4}$ scaling behavior at small $q$ 
usual in coarsening systems, and we also use 
the so-called demixing or inhomogeneous parameter $\psi_n$. 
This is defined by dividing the system in $n^3$ boxes and measuring the
average local excess density
\be
\psi_n= \frac{1}{n^3} 
\sum_{i=1}^{n^3} \left[\rho_i - \langle \rho \rangle \right] \, , \nonumber
\ee
where $\rho_i$ is the local density in the $i$-th box, and $\langle \rho \rangle=N/V=6\phi/(\pi \sigma^3)$ is the average system density. For a homogeneous
phase the demixing parameter is zero, but it grows when density inhomogeneities appear, as those characterizing liquid-gas coexistence.
Using these observables we detect both nucleation and spinodal regimes depending on the quench depth, see Figs.~\ref{snap}-b and \ref{snap}-c, 
respectively.
 
For $\phi>0.05$ and moderate values of $R$ 
a broad equilibrium gel region exists 
between the sol phase at low-$R$ and phase separation at large-$R$, 
see right panel in Fig.~\ref{phase}.  In the gel phase, 
a system-spanning homogeneous cluster of polymer-connected droplets emerges, 
Fig.~\ref{snap}-d, which endows the fluid with 
viscoelastic properties. The observation of an equilibrium gel regime 
in the phase diagram results from the 
independence of the lifetime and the strength of the polymer 
bonds linking the droplets, and makes this model very 
interesting to test at once different
and usually contradictory theoretical frameworks for the gelation 
transition~\cite{Zacca}. 

\begin{figure}
\includegraphics[height=.25\textheight,clip]{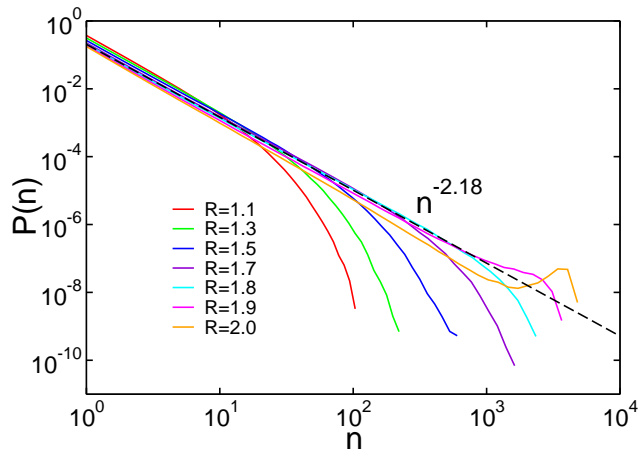}
\caption{\small Distribution of cluster sizes for $N=10^4$, $\phi=0.2$ and increasing values of $R$ across percolation. 
Notice the power-law behavior $P(n)\sim n^{-2.18}$ near percolation (shown as a dashed line), in agreement with random bond percolation, and the peak for 
large cluster sizes above percolation.
}
\label{clust}
\end{figure}

The low-$R$ sol-gel transition coincides in our model with geometric percolation of polymer-connected clusters. The nature of this gelation 
transition is \emph{geometric}, rather than thermodynamic, but 
it cannot be directly detected in experiments with structural probes.
In fact, the structure of the system does not change across the transition, as detected by the structure factor $S(q)$, which remains typical 
of a simple homogeneous fluid, see Fig. \ref{snap}-d. This is also observed in 
experiments~\cite{porte}.  However, despite this apparent homogeneity, the spanning cluster 
is highly fractal near percolation, and becomes thicker deeper in the gel phase. This can be detected in the low wavevector behavior of
$S_{lc}(q)$, the structure factor restricted to droplets belonging to the largest cluster, which exhibit power-law scaling at low-$q$ with an exponent related 
to the cluster fractal dimension. On the other hand, at percolation the distribution of cluster sizes, $P(n)$, shows scaling behavior typical of
the random bond percolation universality class. In Fig.~\ref{clust},
we show $P(n)$ for $\phi=0.2$ and different values of $R$ 
across percolation. As we approach percolation, the cluster distribution develops an algebraic tail with an exponential cutoff which shifts to 
larger and larger values 
as the transition is approached. Above the transition, a peak for large cluster sizes appears 
due to the presence of a system-spanning macroscopic cluster. Near percolation, we find $P(n)\sim n^{-\gamma}$, with $\gamma\approx 2.18$ compatible 
with random bond percolation, as seen in other systems~\cite{napoli,fs1,ema,fs2}. Moreover, the crossover in the scaling of $P(n)$ 
allows us to determine the critical polymer concentration
quite accurately. For $\phi=0.2$, we find this critical value to be
$R_p(\phi)=1.85 (5)$. 
Unfortunately, typical experiments have no access to both $S_{lc}(q)$ 
and $P(n)$ because they cannot discriminate between different clusters, 
so no structural signature of the equilibrium gelation transition is 
actually detected in experiments. 
However, we show below that the dynamics of the system is strongly affected by 
percolation, which can then be used to locate the transition.

The coexistence line ends at a second-order critical point, see Fig. \ref{phase}. The structure of the system becomes 
highly correlated when approaching this liquid-gas phase transition. 
At this point, we find that $S(q)$ develops an algebraic tail with an 
exponent close to $-1.5$ at small $q$. We made no effort in determining
accurately the critical point, but broadly estimated its location
near $\phi_c \approx 0.05$ and $R_c \approx 3.5$. 

\begin{figure}
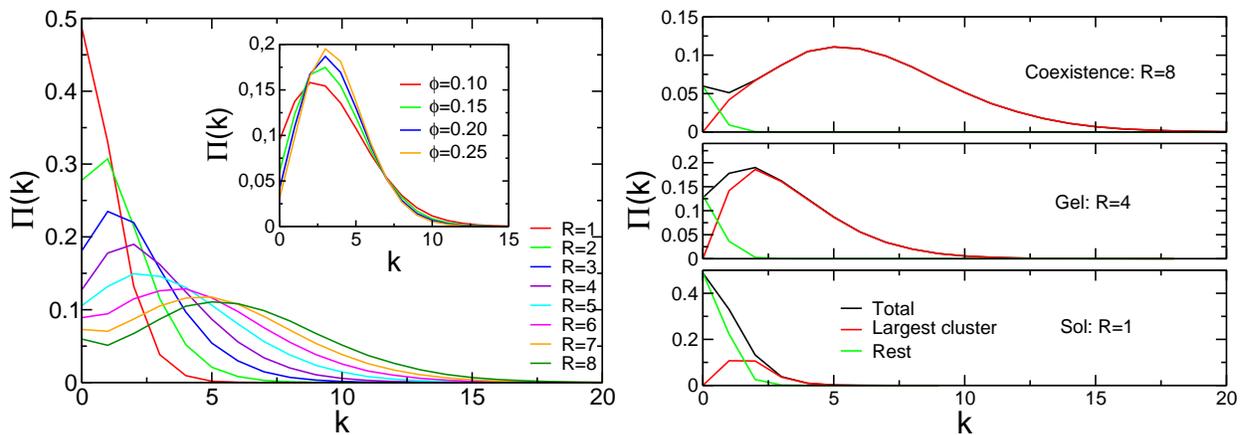

\includegraphics[height=.24\textheight,clip]{degree-distribution-phi_0.1-several-r-and-r_4-several-phi-N_10000.eps}
\includegraphics[height=.24\textheight,clip]{degree-distribution-phi_0.1-several-r-DECOMPOSED-N_10000.eps}
\caption{\small Left: Connectivity distribution $\Pi(k)$ for $N=10^4$, $\phi=0.1$ and 
different values of $R$. Notice that for large $R$ the distribution
becomes bimodal, with a peak at $k=0$, reflecting phase coexistence.
The inset shows $\Pi(k)$ for $R=4$ and increasing 
values of $\phi$.
Right: $\Pi(k)$ and its decomposition for particles which belong to the largest cluster or not, for $\phi=0.1$ and different values 
of $R$ representative of the three phases in Fig. \ref{phase}.}
\label{network}
\end{figure}

We have also studied the statistics of droplet connectivity due to polymer bridges. We define $\Pi(k)$ as the connectivity
distribution, i.e. the probability density for finding a droplet with $k$ linked neighbors. Figure~\ref{network} (left) shows $\Pi(k)$
measured for $\phi=0.1$ and different values of $R$ representative of the three phases in Fig.~\ref{phase}. 
In all cases the connectivity distribution decays exponentially fast 
for large values of $k$, with a typical connectivity scale which 
increases with $R$. One can also detect a crossover from unimodal to bimodal behavior in $\Pi(k)$ as the coexistence line is crossed. In particular, 
for large $R$ the distribution $\Pi(k)$ shows a weak peak at $k=0$ associated to completely disconnected droplets freely diffusing in the low-density phase. 
Such unimodal-bimodal crossover in $\Pi(k)$ is a further evidence 
for phase coexistence.
The inset to the left panel in Fig. \ref{network} shows the $\phi$-dependence of $\Pi(k)$ for a fixed value of $R=4$. Interestingly in the low-$\phi$ regime studied the
connectivity distribution becomes more homogeneous (i.e. peaked) as $\phi$ increases. This might be due to the influence of the liquid-gas critical point. For 
$\phi=0.1$ and $R=4$, which is relatively close to the critical region, large density fluctuations associated to the critical region are still apparent, and this inhomogeneity
in the droplet structure is reflected in a more heterogeneous degree
distribution (see also below). However, increasing $\phi$ moves the system away from 
the critical region, the system becomes 
more homogeneous and so does the degree distribution. 

We can also decompose the connectivity distribution between a contribution 
from the largest cluster present in the system and its complement. 
This is done in the right panel in 
Fig.~\ref{network} for $\phi=0.1$ and different values of 
$R$ representative of the three phases (sol, gel, phase coexistence). 
In all cases we detect a clear asymmetry 
between the connectivity distribution of the largest cluster and the 
rest of the system: most of the highly connected droplets belong to 
the largest cluster. 
This also means that the largest cluster concentrates more polymer 
bridges than the average.
Such asymmetry is most important in the coexistence regime, 
see Figs. \ref{network} (right) and \ref{snap}.b-c, where most bridging polymers 
are concentrated in the high-density droplet phase.

The above discussion concerns relatively low volume fractions, 
$\phi < 0.30$.  Interesting questions arise when pondering over the behavior 
of our model system at higher values of $\phi$.
At higher volume fraction, we anticipate that the 
gel dynamics described in this work will 
compete with the kinetic arrest due to the  
approach of the glass transition usually observed in colloidal
systems at high density.

\section{Slow and Heterogeneous Dynamics}

\begin{figure}[t]
\includegraphics[height=.35\textheight,clip]{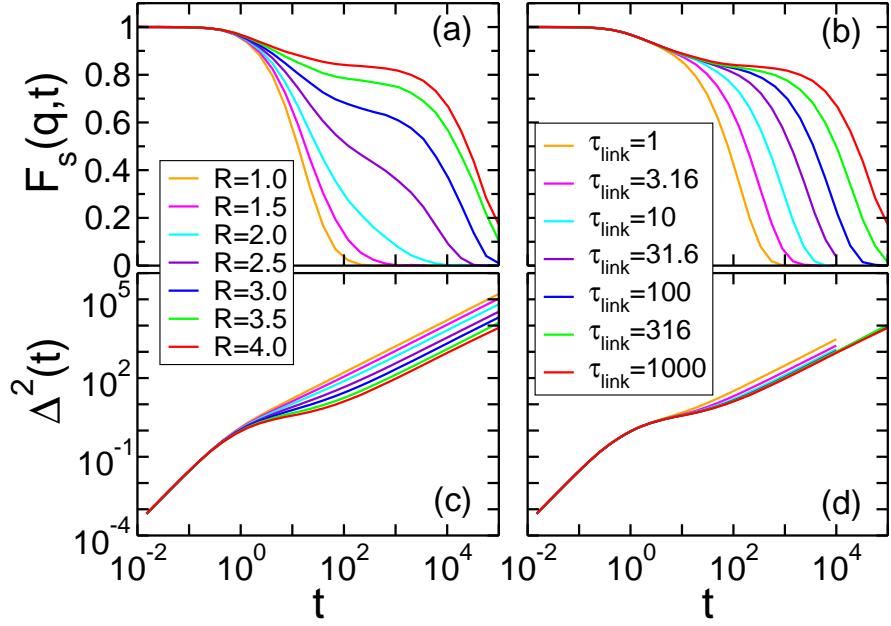}
\caption{\small Self-intermediate scattering function for $q=0.46$ (a, b) and mean-squared
displacement (c, d) for $N=10^3$, $\phi=0.2$. (a, c) present the dynamics for $\tau_{\rm link}=10^3$
and several values of $R$ across percolation ($R_{\rm p}\approx 1.85$ for $\phi=0.2$). Viscoelasticity continuously emerges at percolation. 
(b, d) are for $R=4$ and different values of $\tau_{\rm link}$, which directly controls the long-time decay of $F_s(q,t)$, while 
$\Delta^2(t)$ remains essentially unchanged because its behavior is dominated by fast particles not belonging to the percolating cluster.}
\label{perco}
\vspace{0.8cm}
\end{figure}

In this section we study the dynamics of the equilibrium gel, showing that the gel phase indeed behaves 
dynamically as a soft viscoelastic fluid. We investigate the dynamics by measuring the self-intermediate scattering function
\be
F_s(q,t) =  \left\langle \frac{1}{N} \sum_{j=1}^N \exp[i {\bf q} \cdot ({\bf r}_j(t) - {\bf r}_j(0) )] \right\rangle \, , \nonumber
\ee
and the mean-squared displacement
\be
\Delta^2(t) =  \left\langle \frac{1}{N} 
\sum_{j=1}^N 
| {\bf r}_j(t) - {\bf r}_j(0) |^2 \right\rangle \, . \nonumber
\ee
Figure~\ref{perco}-a shows the evolution of $F_s(q,t)$ from the sol to the gel phase. While relaxation is fast and exponential 
in the sol phase, a slow secondary relaxation emerges at percolation. The final decay time varies little in the gel phase,
but the height of the plateau at intermediate times changes strongly 
across percolation. 
A similar behavior is found for the coherent scattering function, as in experiments~\cite{porte}. 
Physically the plateau reflects the thermal vibrations of an elastic solid on intermediate timescale, while the long-time decay reflects 
the flow of the system: The system is viscoelastic. In Fig.~\ref{perco}-b we show that viscous flow is mostly controlled 
by $\tau_{\rm link}$, the rate for polymer extraction \cite{napoli,fs3}. Flow in this system occurs when the percolating network slowly rearranges 
through polymer moves~\cite{tanaka}. Therefore, gelation corresponds to the continuous emergence, for increasing
polymer concentration, of a plateau in dynamic functions, with an almost constant relaxation timescale,
controlled by the polymer dynamics. Gelation is thus qualitatively different from 
a glass transition where the plateau height remains constant with a dramatic increase of 
relaxation timescales~\cite{debe}, though in both cases no structural signature of the transition is found. 
Coincidence of gelation and percolation, put forward in~\cite{napoli} or
dispelled in~\cite{fs2}, happens whenever long-lived bonds make cluster restructuration
very slow, but does not occur in systems where the bond lifetime is short at percolation~\cite{fs1}.
Notice that the height of the plateau in $F_s(q,t)$ for small $q$ as a function of $R$ can be used as an 
order parameter for the gelation transition.

\begin{figure}[t]
\includegraphics[height=.35\textheight,clip]{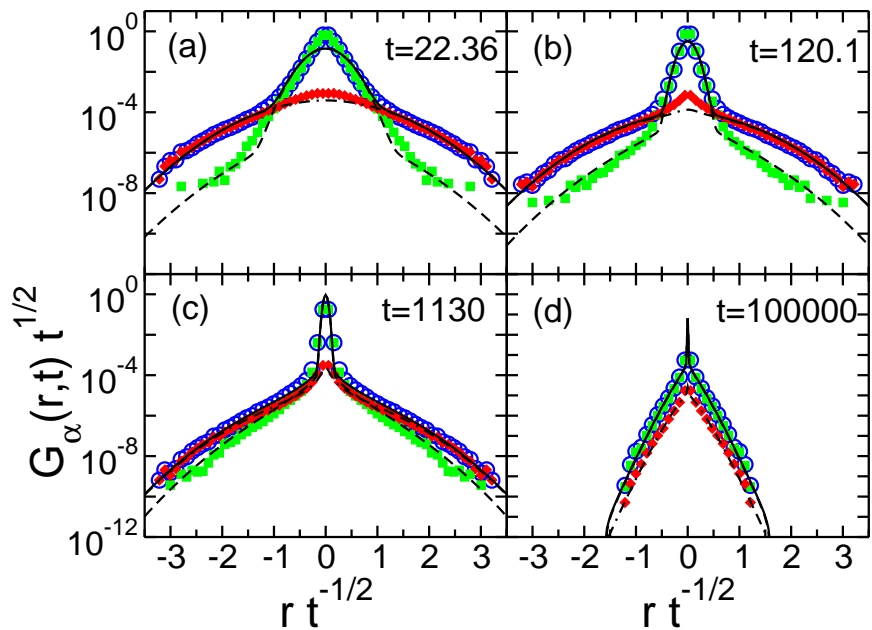}
\caption{\small Distribution of droplet displacements for $\phi=0.2$, $R=4$, 
$\tau_{\rm link}=100$ and different times for all droplets ($\bigcirc$), 
and its decomposition over droplets that
are free ($\diamond$) or arrested ($\Box$) at time $t=0$. 
Lines are from a simple theoretical model based in this decomposition and introduced in Ref. \cite{we}.}
\label{diff}
\end{figure}

Surprisingly, the mean-squared displacements shown in Fig.~\ref{perco}-c
and \ref{perco}-d appear as very poor indicators of the dynamics.
The comparison with the behavior of the self-intermediate scattering function
is in fact quite striking. While the final relaxation timescale, $\tau$, 
in $F_s(q,t)$ scales roughly as $\tau_{\rm link}$, the diffusivity, $D_s$,  extracted from the long-time behavior of 
$\Delta^2(t)\sim 6D_st$ is almost constant. This is reminiscent of the ``decoupling'' phenomenon, or 
``breakdown'' of the Stokes-Einstein relation, reported in supercooled fluids~\cite{debe}. While
``fractional'' breakdown is reported in liquids, $D_s \sim \tau^{-\zeta}$, with $\zeta$ in the range 0.7-0.9 instead of the normal
value $\zeta=1$~\cite{mark}, we find here $\zeta \approx 0$, quite an extreme case of decoupling. Decoupling in gels
was reported in different systems before~\cite{decoupling}.

In supercooled fluids, decoupling phenomena are commonly attributed to the existence of dynamic heterogeneity, that is, the existence
of non-trivial spatio-temporal distributions of mobilities. The analogy is confirmed in Fig.~\ref{diff} where we show distributions 
of droplet displacements, 
\be 
G_s(r,t) = \left\langle \frac{1}{N} \sum_{j=1}^N \delta(|{\bf r}-{\bf r}_i(t)+
{\bf r}_i(0)|) \right\rangle. 
\ee
Clearly, $G_s(r,t)$ 
exhibits a bimodal character suggesting coexistence of slow arrested droplets and fast diffusing droplets.
Qualitatively similar distributions were reported in 
gels~\cite{puertas,colloids,dibble,pinaki2} 
and glasses~\cite{mark,weeks,pinaki}. 
Here, the snapshots in Figs.~\ref{snap}-e and \ref{snap}-f 
suggest an obvious explanation for dynamic heterogeneity. At any given
time, a system-spanning cluster of droplets which behaves as a solid on timescales smaller than $\tau_{\rm link}$
coexists with droplets which can diffuse through this arrested structure. We quantitatively confirm this interpretation in
Fig.~\ref{diff} where $G_s(r,t)$ is decomposed over two families of droplets: $G_s(r,t) = c_A G_A(r,t)+ (1-c_A)G_M(r,t)$, where $A$ and $M$ stand for 
droplets that are Arrested and Mobile at time $t=0$, 
respectively; $c_A$ is the fraction of droplets belonging to the 
percolating cluster. While the central peak in $G_s(r,t)$ is dominated by 
$G_A(r,t)$, the large ``non-Gaussian'' tails are dominated by $G_M(r,t)$.

A simple analytical model can be proposed to describe the dynamic heterogeneity of the equilibrium gel, which incorporates 
the physical idea of the coexistence between a slow, 
percolating cluster of connected droplets
and fast, more freely diffusing droplets, with a dynamic exchange between the two families set by polymer moves. 
Similar physical ideas were qualitatively discussed earlier~\cite{puertas2,decoupling,colloids}, but were, 
however, not exploited within a quantitative model.
The proposed model, described in detailed in Ref.~\cite{we}, 
can be solved analytically in the Fourier-Laplace domain and 
yields quantitative predictions for $G_s(r,t)$ 
in terms of a few free parameters  
(as for instance the concentration $c_A$ of arrested droplets, or the diffusivity of mobile particles) which have a clear physical interpretation
and can be fixed by simple numerical observations. The lines in Fig. \ref{diff}, which agree nicely with measured data, 
are predictions derived from this model. We thus find that the diffusion constant is in fact entirely 
dominated by those droplets which do not contribute to viscoelasticity, and is therefore a poor indicator of the gel dynamics.
These results show that dynamic heterogeneity in gels can be stronger than in supercooled fluids, 
but its origin is also much simpler: The structure of the system is heterogeneous, Fig.~\ref{snap}, while no such static structure 
needs to be invoked in glasses to account for single particle dynamic heterogeneity~\cite{pinaki}.

\section{Conclusions}

In this work we studied a simplified model of a specific material : a microemulsion of oil
droplets in water mixed with telechelic polymers which can form bridges between the droplets.
Although motivated by this material, the model for reversible gelation described here 
sheds light on the microscopic aspects of gelation and the heterogeneous
dynamics of gel-forming systems. 
Our modelling approach neglects the solvent, and uses droplets and polymers as
elementary objects, with a coarse-grained description of polymers as simple
links inducing an effective attraction between the droplets they connect. This
system can be efficiently simulated using a hybrid molecular dynamics / Monte
Carlo scheme.

The model exhibits different phases as the control parameters, $\phi$ and $R$, are
varied. We find in particular a sol phase with liquid-like structure and dynamics, a coexistence region with
both nucleation and spinodal decomposition regimes ending at a liquid-gas phase transition, 
a gelation line determined by geometric percolation, and an equilibrium gel phase unrelated to phase separation. 
The global structure of the equilibrium gel is homogeneous and no structural
differences with the sol phase are detected. 
Despite this apparent homogeneity, the stress is supported by a
fractal network which endows the system with viscoelastic properties. 
Moreover, cluster statistics and geometry at gelation are compatible with the random bond
percolation universality class. 

We also show that, as opposed to structure, dynamics is strongly
affected by gelation, where a slow secondary relaxation appears in time
correlation functions related with the polymer rearrangement timescale, $\tau_{\text{link}}$. In this
way, dynamical (but no structural) changes can be used to locate the gelation transition. 
The equilibrium gel dynamics is highly heterogeneous as a result of the
presence of particle families with very different mobilities.
In fact, at any given time, a system-spanning cluster of droplets which behaves as a solid on timescales smaller than $\tau_{\rm link}$
coexists with droplets which can more freely diffuse through this arrested
structure. These observation motivated a simple yet accurate analytic modelling 
of dynamic heterogeneity, which is generally applicable to gels \cite{we}. Although slow 
and heterogeneous dynamics are reminiscent of the 
physics of supercooled fluids,
we discussed several qualitative differences between gels and glasses.

In this paper we have discussed how gelation plays out at low volume fractions. At high volume fraction,
crowding will result in glassy dynamics, and 
interesting questions arise when gel dynamics
competes with glassiness. We are currently investigating this issue.

\begin{acknowledgments}
We thank G. Porte, C. Ligoure, and S. Mora for motivating discussions, 
and ANR DynHet and Tsanet, MEyC  No. FIS2005-00791, 
and Universidad de Granada for financial support.
\end{acknowledgments}

\end{document}